%% file: main.tex
\documentclass[%
 reprint,
 amsmath,amssymb,
 aps,longbibliography,
 prd
]{revtex4-1}

\usepackage{amsthm}
\usepackage{lmodern}
\usepackage{graphicx}
\usepackage{dcolumn}
\usepackage{bm}
\usepackage{hyperref}
\hypersetup{linktocpage,colorlinks,citecolor={blue},pdfdisplaydoctitle=true,pdfpagemode=UseOutlines,bookmarksnumbered=true}
\usepackage{mathrsfs,dsfont}
\usepackage{color}
\usepackage{microtype}

\definecolor{cbl}{rgb}{0,0,1}
\definecolor{darkblue}{rgb}{0.0,0.0,0.0}

\definecolor{crd}{rgb}{0.0,0,0}

\newcommand{\bra}[1]{\langle #1 |} 
\newcommand{\ket}[1]{| #1 \rangle } 
\newcommand{\upd}{\mathrm{d}}

\newcommand{\eg}[0]{\textit{e.g.} }

\newcommand\e{\mathrm{e}}

\DeclareMathOperator*{\argmin}{argmin} 
 
\newcommand{\hpi}{\hat{\pi}}
\newcommand{\hpsi}{\hat{\psi}}
\newcommand{\hphi}{\hat{\phi}}
\newcommand{\hb}{\hat{b}}
\newcommand{\re}{\mathrm{Re}}
\newcommand{\im}{\mathrm{Im}}

\begin{document}
\title{Gaussian Continuous Tensor Network States for Simple Bosonic Field Theories}
\author{Teresa D. Karanikolaou}
\author{Patrick Emonts}
\author{Antoine Tilloy}
\email{antoine.tilloy@mpq.mpg.de}
\affiliation{Max-Planck-Institut f\"ur Quantenoptik, Hans-Kopfermann-Stra{\ss}e 1, 85748 Garching, Germany}
\affiliation{Munich Center for Quantum Science and Technology (MCQST), Schellingstr. 4, D-80799 München}
\date{\today}

\begin{abstract}
\noindent
Tensor networks states allow to find the low energy states of local lattice Hamiltonians through variational optimization. Recently, a construction of such states in the continuum was put forward, providing a first step towards the goal of solving quantum field theories (QFTs) variationally. However, the proposed manifold of continuous tensor network states (CTNSs) is difficult to study in full generality, because the expectation values of local observables cannot be computed analytically. In this paper, we study a tractable subclass of CTNSs, the Gaussian CTNSs (GCTNSs), and benchmark them on simple quadratic and quartic bosonic QFT Hamiltonians. We show that GCTNSs provide arbitrarily accurate approximations to the ground states of quadratic Hamiltonians, and decent estimates for quartic ones at weak coupling. Since they capture the short distance behavior of the theories we consider exactly, GCTNSs even allow to renormalize away simple divergences variationally. In the end, our study makes it plausible that CTNSs are indeed a good manifold to approximate the low energy states of QFTs.
\end{abstract}

\maketitle

\input{_Introduction.tex}
\input{_CTNS.tex}

\input{_Quadratic.tex}
\input{_Quartic.tex}

\input{_Discussion.tex}

\begin{acknowledgments}
\noindent
We are grateful to Tommaso Guaita, Cl\'ement Delcamp, and Ignacio Cirac for helpful discussions.
P.E. acknowledges support from the International Max-Planck Research School
for Quantum Science and Technology (IMPRS-QST) as well as support by the EU-QUANTERA
project QTFLAG (BMBF grant No. 13N14780). T.D.K. acknowledges support from DAAD.
\end{acknowledgments}

\appendix

\input{_Appendix.tex}

\bibliography{main}

\end{document}

%% file: _Introduction.tex
\section{Introduction}

\noindent
Quantum field theories (QFTs) are difficult to solve out of the perturbative regime with deterministic techniques. Apart from lattice Monte-Carlo algorithms~\cite{wilson1974,creutz1979,bhanot1980,flag2014}, an option would be to solve strongly coupled QFTs variationally. In a nutshell, this would mean guessing a ``good'' manifold $\mathcal{M}$ of states $\ket{\psi_\nu}$ described by a manageable number of parameters $\nu$, minimize the energy $\bra{\psi_\nu}H\ket{\psi_\nu}$ over this class $\mathcal{M}$, and hope that the answer is close enough to the real ground state $\ket{0}$. As was noted by Feynman already \cite{feynman1988}, finding such a good manifold for typical QFTs is a highly non-trivial task. In particular, apart from simple Gaussian states such as free ground states, it seems impossible to have a sparsely parameterized state with easily computable local observables $\bra{\psi_\nu}\mathcal{O}(x_1)\cdots \mathcal{O}(x_n)\ket{\psi_\nu}$ while keeping an extensive ansatz -- the latter requirement excluding \eg simple expansions in the particle number basis.

On the lattice, the situation has proved more favorable in the last two decades. Tensor network states (TNSs) have essentially provided what one was looking for: a sparse and extensive parameterization of many physically relevant many-body quantum states \cite{cirac2009-rev,evenbly2014-rev,molnar2015}. In this approach, the quantum state is obtained from low-rank tensors, contracted along the links of a network. In the translation-invariant case, all tensors are identical, making parameter economy and extensivity manifest. Tensor networks have proved successful numerically in $d=1$ space dimension, with Matrix Product States (MPSs) \cite{fannes1992}, which are at the root of the earlier density matrix renormalization group (DMRG) \cite{white1992,white1993,schollwock2005}, and more recently in $d\geq 2$ with projected entangled pair states (PEPSs) \cite{verstraete2004,vanderstraeten2016,nielsen2017,vanderstraeten2019}. As is often the case, a computationally successful method brings theoretical insights, and tensor network states have allowed a succinct classification of symmetry protected \cite{pollmann2010,chen2011,schuch2011,chen2013} and topological phases of matter \cite{schuch2010,bultinck2017}.

Given their undeniable success on the lattice, it is tempting to try to bring tensor networks to the continuum. This can in principle be done in two ways: by discretizing continuum theories to the lattice, where powerful techniques can be applied more or less out of the box, or by bringing the tensor network toolbox itself to the continuum. While the efficiency of the first approach is so far unmatched \cite{milsted2013,vanhecke2019,delcamp2020}, we here wish to explore the second, longer term option. It turns out that bringing tensor network states to the continuum can be done rather straightforwardly in $d=1$ space dimension, with the so called continuous matrix product states (CMPSs) \cite{verstraete2010} which have been applied successfully to a few QFTs \cite{haegeman2010,stojevic2015,rincon2015}. Going to $d\geq 2$ space dimensions has proved more difficult. Recently, a candidate higher dimensional continuous tensor network state (CTNS) was presented \cite{tilloy2019}. It is obtained as a continuum limit of a lattice tensor network state, and many of the properties of the discrete follow through to the continuum. However, the efficiency of this candidate at approximating low energy states of QFTs has not been demonstrated yet, primarily because carrying computations in the general case is substantially more difficult than with CMPSs. 

Our objective here is to assess the soundness of CTNSs for the approximation of ground states of simple QFTs. To this end, we will restrict ourselves to an easily manageable subclass, the Gaussian CTNSs (GCTNSs). Naturally, this class is very restrictive, and not dense in the space of low energy states of interacting QFTs. It can approximate with arbitrary precision  only the ground states of Hamiltonians quadratic in creation and annihilation operators. But understanding this class is a necessary sanity check: if GCTNSs do not even work in this simple setup, they should probably be abandoned right away. Apart from the assessment of the promise of CTNSs in general, the study of GCTNSs can also provide an economical approximation of physically relevant Gaussian states, which in theory require an infinite number of parameters (a full continuous two-point correlation function) to be defined in the context of QFTs. 

We start our exploration by recalling the definition of CTNSs, characterize their Gaussian submanifold, and give basic computational tools in Sec. \ref{sec:CTNS}. We then apply GCTNSs in Sec. \ref{sec:quadratic} to the task of finding the ground state of a simple quadratic Hamiltonian in $d=1$ and then $d=2$ space dimensions. The higher dimensional setting comes with subtleties related to the infinite energy density of the ground states. Finally, we briefly study in Sec. \ref{sec:quartic} a true interacting model in $d=1$  space dimension, to show how the GCTNS can be efficient in some regimes even if the model under investigation is not quadratic.

%% file: _CTNS.tex
\section{Continuous tensor network states}\label{sec:CTNS}
\noindent
\emph{In this section we define continuous tensor network states (CTNSs), explain how local observables can be computed with them, and introduce a Gaussian subset, the Gaussian CTNSs (GCTNSs) which are analytically tractable.}

\subsection{Definition}

\noindent
A CTNS is a quantum state $\ket{V,\alpha}$, formally belonging to the Fock space $\mathcal{F}[L^2(\mathbb{R}^d)]$, and defined by the functional integral \cite{tilloy2019}:
\begin{equation}\label{eq:pathintegral}
\begin{split}
    \ket{V,\alpha} := 
    &\int\! \mathcal{D} \phi \, \exp\bigg\{\!\!-\!\!\int \upd^d x \; \frac{1}{2}\|\nabla \phi(x)\|^2 \\
    &\hskip1.5cm+ V[\phi(x)] - \alpha[\phi(x)] \, \hpsi^\dagger(x)\bigg\} \ket{\mathsf{vac}},
\end{split}
\end{equation}
where $\ket{\mathsf{vac}}$ is the ``physical'' Fock vacuum state, $(\hpsi ^\dagger,\hpsi)$ are the canonical bosonic creation-annihilation operators on this Fock space, $[\hpsi(x),\hpsi^\dagger(y)] = \delta^d(x-y)$. The ``auxiliary'' field $\phi$ integrated over has $D$ components, $\phi=[\phi_k]_{k=1}^D$, and  $\|\nabla\phi\|^2:= \sum_k \nabla \phi_k \cdot \nabla \phi_k$. This number $D$ is the field bond dimension or simply bond dimension and is the continuous analog of the bond dimension for discrete tensor network states. We have restricted ourselves to the translation invariant case and taken the thermodynamic limit, which spares us the discussion of what happens at the boundaries. Our objective in this paper is to use this quantum state \eqref{eq:pathintegral} as an ansatz for the ground state of a QFT Hamiltonian of interest.

Some quick comments are in order. The state is not normalized, and not all choices of functions $V$ and $\alpha$ even yield a state at all (for example if $V[\phi]=-\phi^2$). We just assume that we choose functions such that the functional integral in \eqref{eq:pathintegral} at least formally makes sense. 

The state is parameterized by two (complex) functions, which suggests that there is an infinite number of parameters even for a number $D$ of auxiliary fields fixed. In practice, one could expand both functions as polynomials in the fields:
\begin{align}
    V[\phi] &= V^{(0)} + V^{(1)}_j \phi_j + V^{(2)}_{jk} \phi_j\phi_k + V^{(3)}_{jk\ell} \phi_j\phi_k \phi_\ell +\dots \, , \nonumber \\
    \alpha[\phi] &= \alpha^{(0)} + \alpha^{(1)}_j \phi_j + \alpha^{(2)}_{jk} \phi_j\phi_k + \alpha^{(3)}_{jk\ell} \phi_j\phi_k \phi_\ell +\dots \, . \nonumber
\end{align}
The maximum degrees $\kappa_V,\kappa_\alpha$ of these two expansions, together with $D$, then give a measure of the expressiveness of the class of states considered. Formally, the coefficients in the expansion are also tensors, and so we recover the simple idea that a tensor network state should associate a quantum state to a few elementary low-rank tensors.

Finally, we may try to give some intuition of the connection between this CTNS ansatz \eqref{eq:pathintegral} and discrete TNSs, for the reader already familiar with the latter. A tensor network state is obtained by taking a product of elementary tensors and contracting a fraction of their indices (the bond indices) along the edges of a lattice. For CTNSs, the equivalent of the product of tensors is the exponential of the integral, the equivalent of the contraction of discrete indices is a product of integrals over auxiliary fields, which becomes a functional integral in the limit \cite{tilloy2019}. The gradient square term in \eqref{eq:pathintegral} comes from the fact that the tensors are connected to their nearest neighbors. In this paper, understanding the derivation of CTNSs as the continuum limit of TNSs is not needed, since we will directly test the validity of CTNSs in the continuum, without relying on a discretization.

\subsection{Generating functional}

\noindent
To compute expectation values of local observables on a CTNS, the most straightforward method is to introduce the generating functional $\mathcal{Z}_{j',j}$ for the normal ordered correlation functions:
\begin{equation}\label{eq:generatingfunction}
\mathcal{Z}_{j',j}=\frac{\bra{V,\alpha} \exp\left(\int j'\, \hpsi^\dagger \right) \exp\left(\int j\, \hpsi  \right) \ket{V,\alpha}}{\langle V,\alpha|V,\alpha\rangle} \,.
\end{equation}
For example, it can be used to compute the simple two-point function
\begin{align}
\langle\hpsi^\dagger(x) \hpsi(y)\rangle_{V,\alpha}:=& \frac{\langle V,\alpha | \hpsi^\dagger(x) \psi(y)| V,\alpha\rangle}{\langle V,\alpha | V,\alpha\rangle} \nonumber \\
=& \frac{\delta}{\delta j'(x)}\frac{\delta}{\delta j(y)} \mathcal{Z}_{j',j} \bigg|_{j,j'=0}\,. \label{eq:correlexamplebis}
\end{align}
Using the Baker-Campbell-Hausdorff formula to commute the two exponentials in \eqref{eq:generatingfunction} and then using the formula for the overlap of unnormalized field coherent states, one obtains \cite{tilloy2019}:
\begin{align}\label{eq:generatingfunctionintegral}
   \mathcal{Z}_{j',j} &=\frac{1}{\mathcal{N}}\int \mathcal{D}\phi\mathcal{D}\phi' \exp\bigg\{- \int \frac{\|\nabla\phi\|^2+\|\nabla\phi'\|^2}{2} \\
   &+V[\phi] +V^*[\phi'] -\alpha^*[\phi']\alpha[\phi] - j\alpha[\phi] - j' \alpha^*[\phi']\bigg\} \,.\nonumber
\end{align}
It is important to note that powers of the field in the expansion of $\alpha$ come multiplied and connect together the two auxiliary fields coming from bra and ket, as in a Schwinger-Keldysh functional integral. In general, if arbitrary powers of the field appear, the functional integral \eqref{eq:generatingfunctionintegral} might be diverging. Assuming that  the divergences can be properly substracted, then actually \emph{computing} correlation functions remains difficult non-perturbatively. Apart from Monte-Carlo techniques, a boundary CMPS method was suggested in \cite{tilloy2019} but we will not explore this further here.

\subsection{Gaussian subset}

\noindent 
The functional integral in \eqref{eq:generatingfunctionintegral} can be computed exactly if the expansions of $V$ and $\alpha$ are truncated to quadratic and linear order respectively \footnote{Beware of the factor of $2$ difference in the definition of $V^{(2)}$ compared to \cite{tilloy2019}}:
\begin{align}
    V[\phi] &= V^{(0)} + V^{(1)}_j \phi_j + V^{(2)}_{jk} \phi_j\phi_k \, , \nonumber\\
    \alpha[\phi] &= \alpha^{(0)} + \alpha^{(1)}_j \phi_j \, .\nonumber
\end{align}
We call states defined by such restricted $V$ and $\alpha$ Gaussian CTNS. Note that these states are Gaussian in the usual sense. Indeed, one can carry the Gaussian integral directly in the state definition \eqref{eq:pathintegral} to get:
\begin{equation}
    \ket{V,\alpha} = \exp\left\{ \int \hpsi^\dagger G\,  \hpsi^\dagger + \beta\, \psi^\dagger\right\}\ket{\textsf{vac}} \, ,
\end{equation}
where
\begin{align}
    G&= \alpha^{(1)\, T}  \left(-\frac{\nabla^2}{2}\mathds{1} + V^{(2)}\right)^{-1} \, \alpha^{(1)}\\
    \beta&=\alpha^{(0)} - \frac{1}{2} \bigg[V^{(1)\, T} \left(-\frac{\nabla^2}{2}\mathds{1} + V^{(2)}\right)^{-1} \alpha^{(1)} \nonumber \\
    &\hskip1.63cm + \alpha^{(1)\, T}\left(-\frac{\nabla^2}{2}\mathds{1} + V^{(2)}\right)^{-1}V^{(1)}\bigg]. \label{eq:beta}
\end{align}
This expression allows to spot a lot of redundancy in the parameterization. The first and simplest observation is that $V^{(0)}$ does not appear because it merely changes the state normalization which we do not keep track of. Second, we notice that the second term in \eqref{eq:beta} can be incorporated into $\alpha^{(0)}$, and thus we may fix $V^{(1)}=0$ without lack of generality. This is quite intuitive: giving the auxiliary field a non-zero expectation value can be compensated by a constant source. Finally, under the mild assumption that $V^{(2)}$ is diagonalizable $V^{(2)}=U^{-1} (M/2) U$, we have a straightforward rewriting:
\begin{equation}
    G= \frac{1}{2}\sum_{\ell} \underset{:= A_\ell}{\underbrace{\bigg[\sum_{jk} \alpha^{(1)}_j U_{j \ell} U^{-1}_{\ell k}  \alpha_k\bigg]}} \bigg( -\nabla^2 + M_{\ell\ell} \bigg)^{-1}.
\end{equation}
This expression could be obtained directly by taking $V^{(2)}=M/2$ diagonal and $\alpha^{(1)}_k$ a complex square root of $A_k$. Thus, without lack of generality, we can now assume that we have diagonal ``mass'' matrix $M:=\text{diag}(m_1,\dots,m_D)$ for the auxiliary field. In the end, a GCTNS is simply parameterized by two complex vectors $\alpha^{(1)}$ and $m$, and a scalar $\alpha^{(0)}$, that is $2 D+1$ complex parameters.

We may now go back to the computation of the generating functional \eqref{eq:generatingfunctionintegral}. Carrying out the Gaussian integral yields:
\begin{equation}
\label{eq:generating_gaussian}
 \mathcal{Z}_{j',j} = \exp \bigg( \int  \frac{1}{2}\, J(j,j')^T \, K \, J(j,j') + j\alpha^{(0)} + j'\alpha^{(0) *}\bigg), \nonumber
\end{equation}
where the operator $K$ fulfills
\begin{equation}
    \left(\begin{array}{cc}
-\nabla^2+M & -\alpha^{(1)}\alpha^{(1)* T}\\
-\alpha^{(1)*}\alpha^{(1) T}  & -\nabla^2+M^{*}
\end{array}\right)K(x,y) = \mathds{1} \, \delta(x-y)\, , \nonumber
\end{equation}
and $J(j,j')^T=(\alpha^{(1)}[j+ \alpha^{(0) *}] ,\alpha^{(1)*}[j'+ \alpha^{(0) }])$. Because of translation invariance $K(x,y)=K(x-y)$, and it is convenient to go to Fourier space:
\begin{equation}
    K(x-y)= \int \frac{\upd^d p}{(2\pi)^d} K(p) \, \e^{ip(x-y)}
\end{equation}
which yields $K(p) = \left(p^2 \, \mathds{1} + W\right)^{-1}$ with
\begin{equation}\label{eq:W}
W=\left( \begin{array}{cc}
M & -\alpha^{(1)} \alpha^{(1)* T}\\
-\alpha^{(1)*}\alpha^{(1)T}  & M^{*}
\end{array} \right).
\end{equation}
With this, we can compute various expectation values of the state, for example the two-point functions using \eqref{eq:correlexamplebis}.

\subsection{Variational optimization}

\noindent
We now summarize the strategy to variationally optimize GCTNSs in practice. In what follows, we will study models specified by a local bosonic Hamiltonian:
\begin{equation}
    H = \int_{\mathbb{R}^d}\upd^d x \;  h(\hpsi^\dagger,\hpsi)(x) \, ,
\end{equation}
where $h(\hpsi^\dagger,\hpsi)(x)$ contains product of the operators $\hpsi,\hpsi^\dagger$ and its derivatives. For a GCTNS $\ket{V,\alpha}$ we introduce the associated energy density
\begin{equation}
    \langle h\rangle_{V,\alpha}:=\frac{\bra{V,\alpha} h(\hpsi^\dagger,\hpsi)\ket{V,\alpha}}{\bra{V,\alpha} V,\alpha\rangle}\, .
\end{equation}
Our objective is to minimize it to find an approximation to the ground state $\ket{0}$ and an upper bound to the ground energy density $\mathbf{e}_0$:
\begin{align}
    \ket{0}&\simeq \ket{V,\alpha} = \argmin \, \langle h\rangle_{V,\alpha} \, ,\\
 \mathbf{e}_0 &\leq \min_{V,\alpha} \, \langle h\rangle_{V,\alpha}  \; .
\end{align}
To carry out the minimization, one needs to be able to compute $\langle h\rangle_{V,\alpha}$, which reduces to the computation of a sum of correlation functions of $\hpsi,\hpsi^\dagger$, which we know how to compute in general from the generating functional (see Appendix \ref{appendix:correl}).

Whether we use simple gradient descent or more advanced optimization algorithm like Broyden–Fletcher–Goldfarb–Shanno (BFGS)~\cite{numerical2007}, we also need the gradient of $\langle h\rangle_{V,\alpha}$ with respect to the $2D+1$ complex coefficients parameterizing the state. Since we have explicit expressions for all correlation functions, this presents no fundamental difficulty and is done in appendix \ref{appendix:correl}.

%% file: _Quadratic.tex
\section{A quadratic model in \texorpdfstring{$1$}{1} and \texorpdfstring{$2$}{2} space dimensions} \label{sec:quadratic}

\noindent
\emph{In this section, we present a simple quadratic, thus exactly solvable, Hamiltonian and approximate its ground state with our GCTNS ansatz.}

\subsection{The model} \label{sec:modeldefinition}

\noindent
We first consider a model with a Hamiltonian quadratic in creation and annihilation operators
\begin{equation}\label{eq:quadH}
    H = \int_{\mathbb{R}^d}\nabla \hpsi^\dagger\nabla\hpsi + \mu \, \hpsi^\dagger\hpsi + \lambda \left[\hpsi^\dagger\hpsi^\dagger + \hpsi\,\hpsi \right]\, ,
\end{equation}
and that thus has a Gaussian ground state.
In fact, for a single species of spinless bosons and the usual non-relativistic kinetic term, it is essentially the most general one can write. Such a Hamiltonian can typically be obtained as the mean field approximation of a weakly interacting Bose gas, but we take it as an exact starting point here. Another instructive way to interpret this Hamiltonian is to see it as the regularized Hamiltonian of the relativistic free boson \cite{stojevic2015}:
\begin{equation}
    H^\Lambda_\text{fb}=\frac{1}{2}\int_{\mathbb{R}^d} \hpi^2+ (\nabla \hphi)^2 + m^2 \hphi^2 + \underset{ \text{regulator}}{\underbrace{\frac{1}{\Lambda^2} \left(\nabla \hpi\right)^2}},
\end{equation}
where $\hpi,\hphi$ are the traditional canonically conjugate fields $[\hphi(x),\hpi(y)]=i\delta^{d}(x-y)$, and $\Lambda$ is a non-relativistic momentum cutoff.
This Hamiltonian $H^\Lambda_\text{fb}$ reduces to \eqref{eq:quadH} with the field mapping
\begin{align}
    \hphi &= \sqrt{\frac{1}{2\Lambda}}(\hpsi+\hpsi^\dagger)\\
    \hpi&= \sqrt{\frac{\Lambda}{2}}\; (\hpsi-\hpsi^\dagger)
\end{align}
and the parameters
\begin{align}
    \mu&=\frac{\Lambda^2 + m^2}{2}\, ,\\
    \lambda&=\frac{\Lambda^2-m^2}{4}\,.
\end{align}
Closing the gap, which happens when $\lambda/\mu \rightarrow f_c=1/2$, is equivalent to lifting the non-relativistic regulator ($m\ll \Lambda$).

This model is exactly solvable and one finds (see appendix \ref{appendix:exact}) that its ground state energy density is
\begin{align}\label{eq:exactedensity}
    \mathbf{e}_0= \frac{1}{2}\int\frac{\upd^d p}{(2\pi)^d} \,\left[\sqrt{(p^2+\mu)^2 - 4\lambda^2} - (p^2+\mu)\right]\, ,
\end{align}
which is infinite when $d \geq 2$. Consequently, in $d=1$, we will be able to directly optimize the energy, whereas in $d\geq 2$, we will have to renormalize away the divergent part. The corresponding two-point functions can also be computed exactly and we have \eg
\begin{align}\label{eq:exactcorrel}
    \bra{0} \hpsi^\dagger(x) \hpsi(y)\ket{0} &= \int \frac{\upd^d p}{(2\pi)^d}  C_0(p)\e^{ip(x-y)} \\
    \text{with} ~~ C_0(p)&=\frac{1}{2}\left( \frac{p^2+\mu}{\sqrt{(p^2+\mu)^2- 4\lambda^2}} -1\right). \nonumber
\end{align}

\subsection{Variational optimization in \texorpdfstring{$1$}{1} space dimension}

\begin{figure*}
    \centering
    \includegraphics[width=\columnwidth]{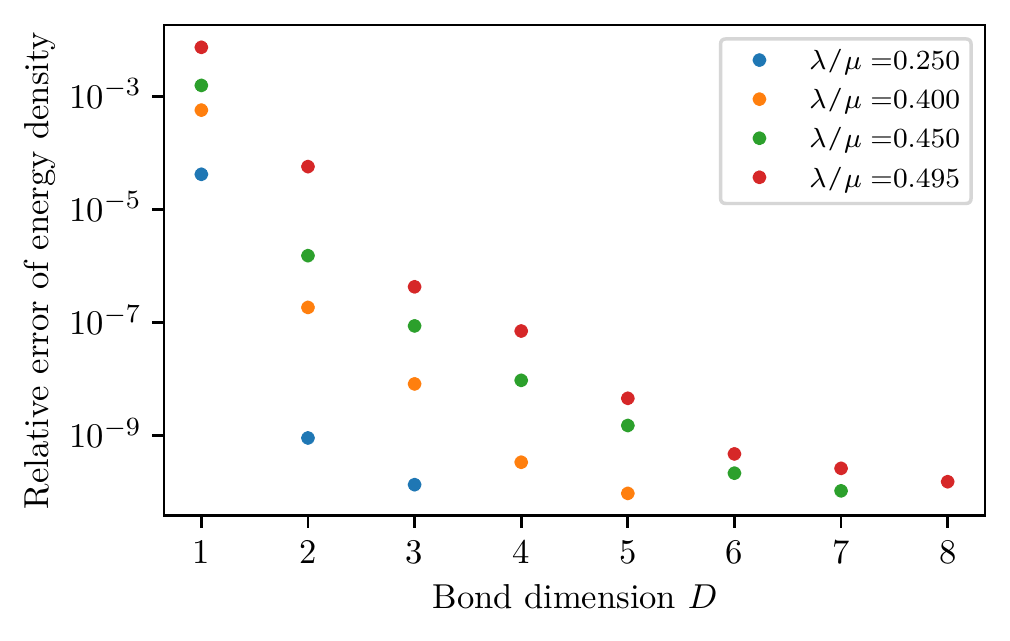}
    \includegraphics[width=\columnwidth]{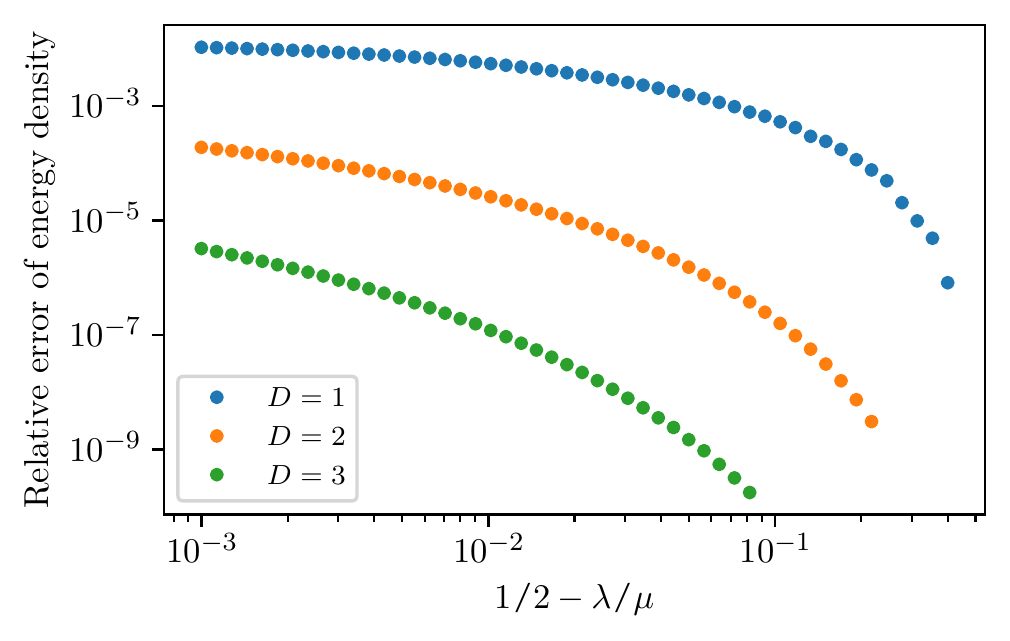}
    \caption{\textbf{Convergence of the energy density in $\mathbf{d=1}$}. Left -- Relative error in the energy density $\langle h\rangle_{V,\alpha}/\mathbf{e}_0-1$ as a function of the field bond dimension D. Right -- Relative error as a function of the distance $1/2-\lambda/\mu$ from the gap closing point for $D=1,2,3$.}
    \label{fig:results_1d_energy}
\end{figure*}

\noindent
To compute the ground state energy with our ansatz, we simply compute the energy density, its gradient with respect to the parameters, and use a standard BFGS solver to find the point yielding the minimal energy. The results are shown in Fig. \ref{fig:results_1d_energy}.

We observe that for parameter values of order $1$ away from the gap closing (say $f=\lambda/\mu=0.25=f_c/2$), the convergence to the exact value is extremely fast in $D$ -- to the point that it is difficult to probe large values of $D$ because of machine precision issues. As we get closer to the gap closing, the convergence becomes slower, but moderate values of $D$ still give accurate values, even for $\lambda/\mu=0.99 f_c$. This is compatible with the TNS folklore that gapped systems can be precisely approximated with low bond dimension, and that larger values have to be used as we get closer to a critical point.

\begin{figure*}
    \centering
    \includegraphics[width=\columnwidth]{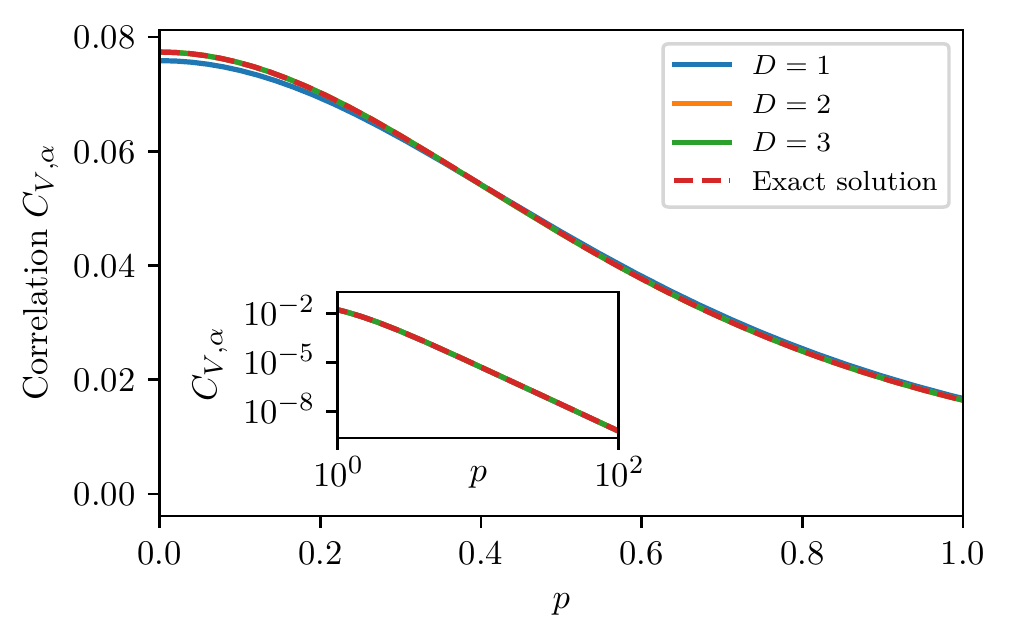}
    \includegraphics[width=\columnwidth]{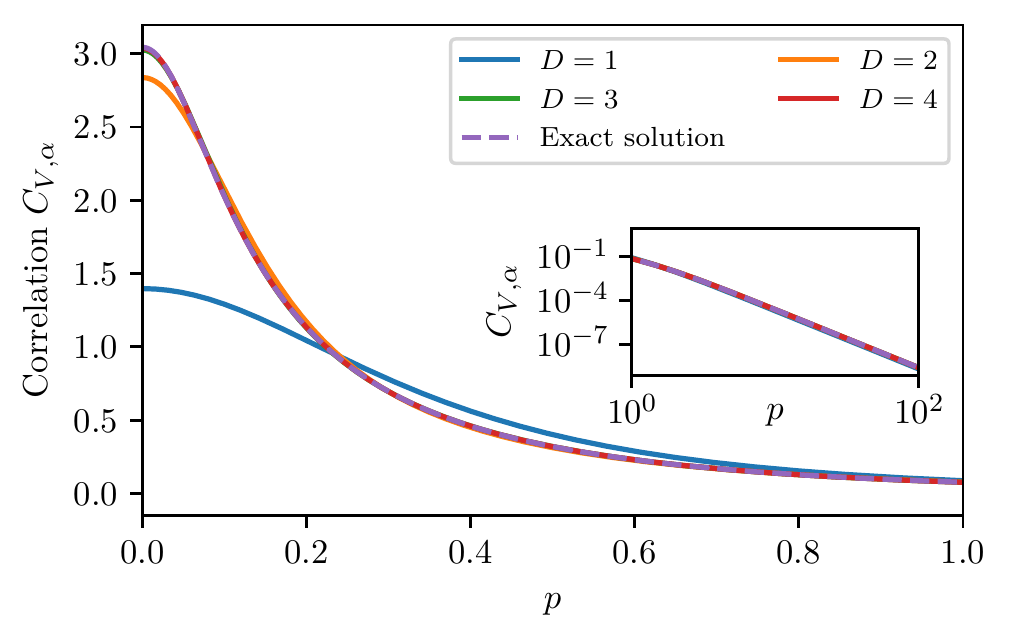}
   
    \caption{\textbf{Convergence of the correlation function in $\mathbf{d=1}$}. Two-point correlation function in momentum space $C_{V,\alpha}(p)$ for $\mu=1$  away from the gap closing for $\lambda = 0.25$ (left) and near the gap closing for $\lambda = 0.495$ (right). The GCTNS correlation function converges uniformly to the exact one as $D$ is increased, but larger values of $D$ are required as the gap closes.}
    \label{fig:correlations_d1}
\end{figure*}

In QFT, one might worry that optimizing the energy does not give a fast convergence of the state itself (summarized by its two-point functions in the Gaussian case). Here, because the theory is regular (or equivalently non-relativistic), this is not the case, and we observe a fast uniform convergence of the two-point function, at least away from the gap closing (see Fig. \ref{fig:correlations_d1}).

\subsection{A theoretical apart\'e}

\noindent
Before we move on to the trickier $d=2$ space dimensions case, it is helpful to understand better the structure of GCTNS correlation functions and compare them to the exact one \eqref{eq:exactcorrel}. Using the expression for the generating functional, it is straightforward to see that $C_{V,\alpha}(p)$, the Fourier transform of $\langle \hpsi^\dagger(x)\hpsi(0)\rangle_{V,\alpha}$, is of the form
\begin{align}\label{eq:CaV}
    C_{V, \alpha} (p) &= \sum_{k=1}^{2D} \frac{a_k}{p^2+w_k},
\end{align}
where $w_k$ are the complex eigenvalues of $W$ defined in \eqref{eq:W} and $a_k$ are complex coefficients (see appendix \ref{appendix:correl} for more detail). Putting all the fractions in \eqref{eq:CaV} on the same denominator shows that $C_{V,\alpha}(p)$ is an even rational function of degree at most $4D$. Clearly, this means that there is no chance to capture $C_0(p)$ exactly for a finite $D$, since it contains a square root. However, using the identity
\begin{equation} \label{eq:expansion1}
    (1-x)^a=\sum_{n=0}^{\infty} \binom{a}{n} \, x^{n},
\end{equation}
we have that
\begin{equation}\label{eq:expansion2}
    C_0(p) = \frac{1}{2}\sum_{n=1}^{+\infty}\binom{-\frac{1}{2}}{n}\, \left[\frac{4\lambda^2}{(p^2+\mu)^2}\right]^n,
\end{equation}
with uniform convergence for all $p$ as long as $\lambda/\mu <1/2$. This is the same structure as a GCTNS correlation function, except that the expansion in rational functions is truncated at order $4D$ in $p$ for GCTNSs.

At short distances, $p\rightarrow + \infty$, only the first term in the expansion matters. It can be reproduced exactly already by a GCTNS with $D=1$, which means the UV behavior of the QFT can be captured by the simplest non-trivial GCTNS. At long distances, $p\simeq0$, the series \eqref{eq:expansion2} is still absolutely convergent with an error decreasing exponentially with the number of terms. Hence, for a GTCNS the error should be dominated by the infrared and at most $\mathcal{O}([2\lambda/\mu]^{2D})$.

Naturally, we perform a variational optimization of the energy and not a perturbative term by term optimization of the two-point function, and as a result the error obtained in practice could scale differently. And indeed, we observe in Fig. \ref{fig:results_1d_energy}, at least for small $D$, that the error decreases faster than naively expected.

\subsection{Variational optimization in \texorpdfstring{$2$}{2} space dimensions and renormalization}

\noindent
In $d=2$ space dimensions, several two point functions of interest diverge when taken at equal points. In particular, the kinetic energy $\nabla \hpsi^\dagger \nabla \hpsi $ and $\hpsi\hpsi + \hpsi^\dagger\hpsi^\dagger$ terms diverge when evaluated on GCTNSs. This can be traced back to the fact that the corresponding momentum integrals (see appendix \ref{appendix:correl}) diverge logarithmically.

This divergence can be renormalized in a way we now explain. First, we introduce a hard momentum cutoff $\Lambda$ such that correlation functions are finite. We then observe that the energy density evaluated on a GCTNS reads:
\begin{equation}
        \langle h \rangle_{V,\alpha}= \langle h\rangle_\text{r}+  \frac{1}{4\pi}\ln(\Lambda^2)\,\langle h\rangle_\text{div}+o(1),
\end{equation}
such that the energy can be split into a regular and log divergent part. For the Hamiltonian \eqref{eq:quadH} we consider, the log divergent part can be evaluated exactly and we find:
\begin{equation}
\begin{split}
  \langle h\rangle_{\text{div}}= \left[\sum_{j=1}^D \alpha^{ 2}_j\right]\left[\sum_{j=1}^D \alpha^{ 2}_j\right]^*  
  +\lambda \sum_{j=1}^D \left(\alpha^{ 2}_j + \alpha^{*2}_j\right),
  \end{split}
\end{equation}
where we used the simplified notation $\alpha_j^{(1)}=\alpha_j$. Importantly, $\langle h\rangle_{\text{div}}$ can be made negative and minimized exactly, yielding the condition:
\begin{equation}\label{eq:condition}
    \sum_{j=1}^D \alpha_j^2 = - \lambda.
\end{equation}
This condition defines a submanifold of ``maximally divergent energy'' GCTNS on which the parameters can be numerically tuned to minimize the remaining finite part $\langle h \rangle_\text{r}$. 

We obtained the condition \eqref{eq:condition} in a variational way, only asking that the energy be minimal and taking the cutoff to infinity. As a welcome surprise, it provides the same divergence of the energy density as the exact solution! Indeed, as can be seen from \eqref{eq:exactedensity}, the latter diverges as $-\lambda^2\ln\Lambda^2/(\pi)$, exactly as for GCTNS on the submanifold defined by \eqref{eq:condition}. So not only can a GCTNS capture the UV behavior of the exact ground state, it captures it \emph{exactly} upon optimization.

In what follows and for comparison, we consider only the renormalized part of the energy density $\langle h \rangle_\text{r} := \lim_{\Lambda\rightarrow \infty} \langle h \rangle - \lambda^2 \ln(\Lambda^2)/(4\pi)$. For the exact solution it gives the ``renormalized'' energy density 
\begin{equation}
   \textbf{e}^R_0 =\int \frac{\upd^2 p}{(2\pi)^2}\;\left(\varepsilon_0(p)+\frac{\lambda^2}{(p^2+\mu)}\right)+\frac{\lambda^2}{4\pi}\ln(\mu),
\end{equation}
which is of course finite and which we can compare to $\langle h_\text{r}\rangle$ since the counter terms used in both cases (the divergent parts) are identical. Again, we insist that this optimization procedure and the associated renormalization of the energy density do not require knowing the exact solution.

\begin{figure}
    \centering
   
    \includegraphics[width=\columnwidth]{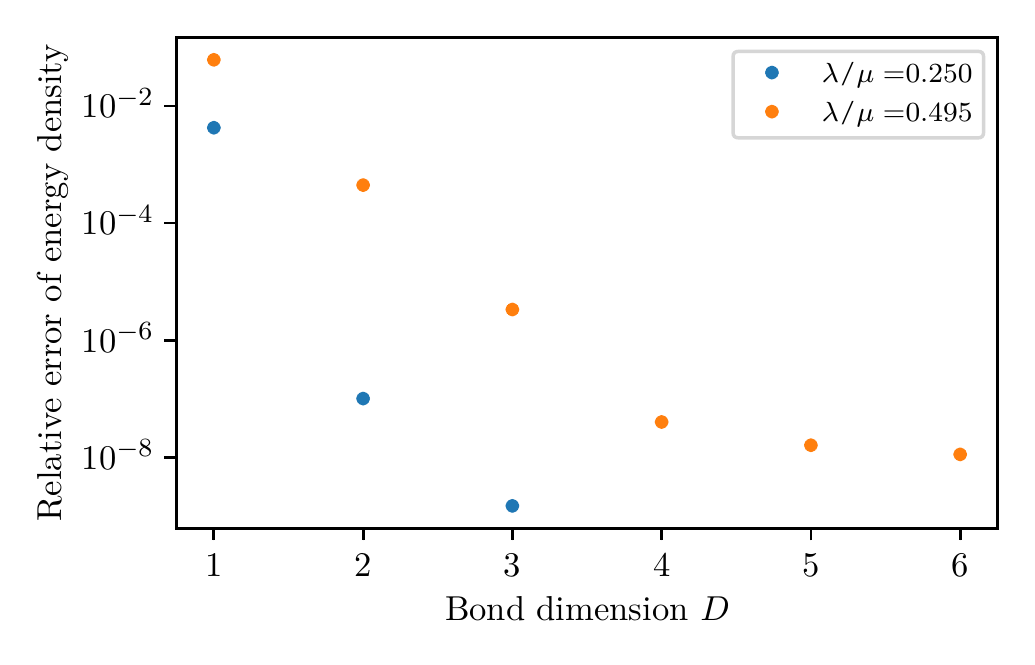}
    \caption{\textbf{Convergence of the energy density in $\mathbf{d=2}$}. Relative error in the energy density as a function of the bond field dimension D for $\lambda/\mu=0.25$ (away from the gap closing) and $\lambda/\mu=0.495$  (close to the gap closing).}
    
    \label{fig:results_2d}
\end{figure}

Results are shown in Fig. \ref{fig:results_2d} and we observe that the convergence for the renormalized energy density and two point function is qualitatively as good as in the $d=1$ case.

\begin{figure*}
    \centering
     \includegraphics[width=\columnwidth]{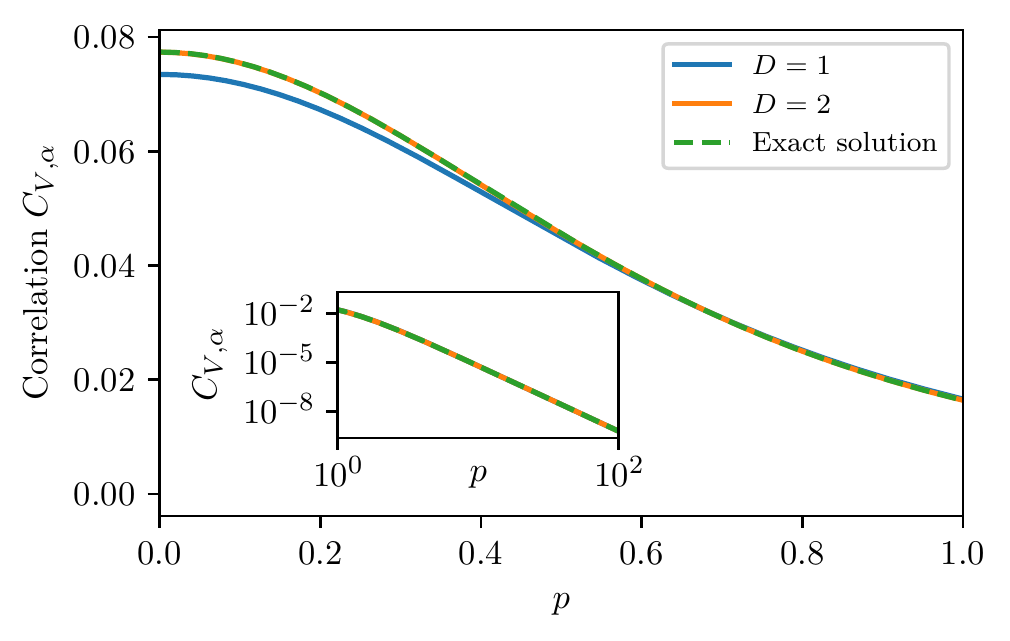}
    \includegraphics[width=\columnwidth]{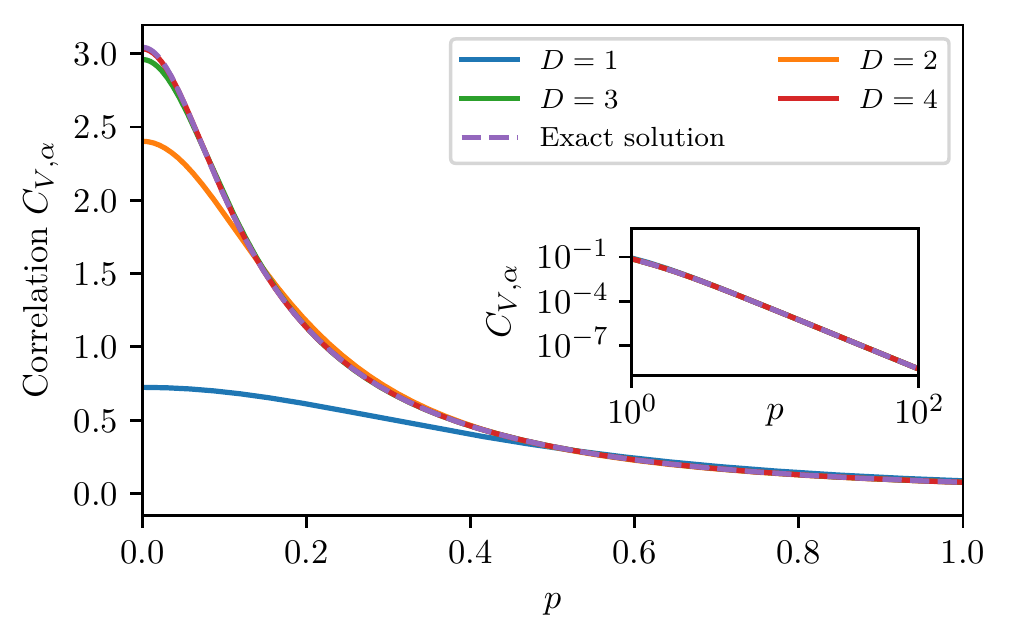}
   \caption{\textbf{Convergence of the correlation function in $\mathbf{d=2}$}. Two-point correlation function in momentum space $C_{V,\alpha}(p)$ for $\mu=1$  away from the gap closing for $\lambda = 0.25$ (left) and near the gap closing for $\lambda = 0.495$ (right). As in $d=1$, the GCTNS correlation function converges uniformly to the exact one as $D$ is increased in two space dimensions.}
    \label{fig:correlations_2d}
\end{figure*}

%% file: _Quartic.tex
\section{A quartic model in \texorpdfstring{$1$}{1} space dimension} \label{sec:quartic}

\noindent
\emph{In this section we study a simple quartic model, the Lieb-Liniger model, that has a non-Gaussian ground state. Consequently, there is no hope to approximate it with arbitrary precision with GCTNS, but we may still capture qualitative features.}

\subsection{Lieb-Liniger model}
\noindent
The Lieb-Liniger model is about the simplest model of interacting bosons in $d=1$ space dimension and is given by the Hamiltonian
\begin{equation} \label{eq:LL_hamiltonian}
    H_\text{LL} = \int_\mathbb{R}  \partial_x \hpsi^\dagger\partial_x\hpsi + c\,  \hpsi^\dagger\hpsi^\dagger\hpsi\hpsi,
\end{equation}
where $c$ is the strength of the coupling. The number of particles is conserved and another important parameter in the model is the particle density $\rho = \langle \hpsi^\dagger \hpsi\rangle$. The physics of the model depends on the adimensional coupling $\gamma = c/\rho$. This model is integrable and with the Bethe Ansatz it is possible to write an exact equation for the energy density in the ground state, which can be solved numerically to essentially arbitrary precision or expanded in a power series at weak and strong coupling \cite{bethe_low,bethe_high}.

The ground state of this model is not a Gaussian state, and as a result a GCTNS cannot approximate it with arbitrarily good precision even for large $D$. However, it is possible for a GCTNS to give reasonable approximation in some regime, which is what we aim to explore here. To this end, we will compare with two other simple approximation techniques: classical solution and mean field. For us, the classical solution is simply what we obtain by minimizing the energy in the space of coherent states, or equivalently GCTNS with $D=0$. The mean-field approximation corresponds to the ground state of a different Hamiltonian, namely the mean field quadratic Hamiltonian of the same model. In appendix \ref{sec:four_point}, we explain how to deal with the quartic terms and how to obtain the mean field Hamiltonian. 

Our analysis can be seen as the continuum analog of the one carried recently for the Bose-Hubbard model \cite{guaita2019}, where a generic Gaussian state approximation was compared with standard classical and mean field solutions. In our case, aside from dealing with the continuum, we have the refinement that we do not use the most general Gaussian states in the first place (which would anyway require infinitely many parameters), but a tower of more and more expressive submanifolds indexed by $D$.

\subsection{Results}

\noindent
In practice, we simply minimize the energy density of the model over GCTNSs of fixed $D$ keeping $\rho=1$ fixed with gradient descent. As GCTNSs are Gaussian states, the expectation value of the quartic term is simply computed with Wick's theorem (see appendix \ref{appendix:correl}), and thus the energy density and its gradient are easily evaluated.

\begin{figure}
    \centering
    \includegraphics[width=\columnwidth]{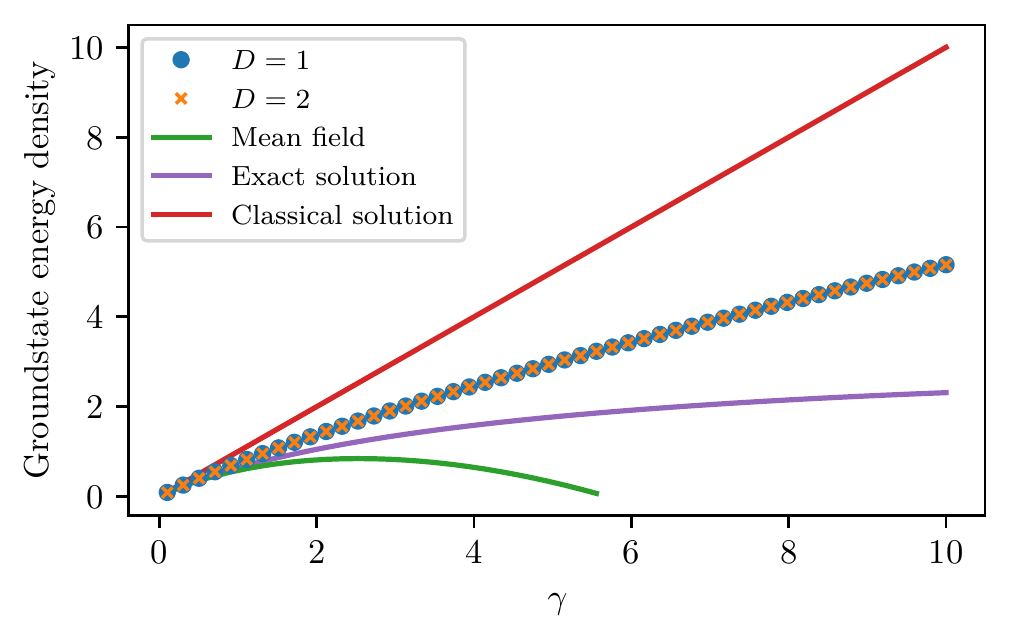}
    \caption{Energy density of the Lieb-Liniger model ground state as a function of the coupling strength $\gamma=c/\rho$.  }
    \label{fig:liebliniger}
\end{figure}

In Fig. \ref{fig:liebliniger}, we can see that the upper bound provided by GCTNS approaches the exact ground energy as the coupling $\gamma$ gets smaller. This is expected: the ground state of a weakly interacting Bose gas becomes Gaussian when the coupling goes to zero. 

What is remarkable is that the simplest GCTNS ansatz for $D=1$ is already sufficient to get all the expressive power of Gaussian states in this case. Almost all the improvement from the classical solution $D=0$ is reached for $D=1$. The refinements obtained with larger $D$ are not necessary in the sense that they bring improvements in the energy density much smaller than the distance between the best Gaussian energy density and the true energy density.
This is rather intuitive: if a Gaussian state is anyway not the exact solution, we do not gain much by getting the absolute best Gaussian state, and a crude approximation of the best Gaussian state can do qualitatively as well.

%% file: _Discussion.tex
\section{Discussion}
\noindent
Let us briefly summarize our results. We were mainly interested in knowing if CTNSs had the right properties to be good trial wave functions for quantum field theories, mirroring the efficiency of their discrete lattice counterparts. To this end, we focused on a subclass of analytically tractable CTNSs, the Gaussian CTNSs, which are also a submanifold of general Gaussian states. Optimizing GCTNSs on a simple non-relativistic quadratic Hamiltonian, we obtained a very good match with the exact solution, both for the energy density and the state itself (parameterized by its two point function). Importantly GCTNSs have the right UV behavior, even for the minimal number of auxiliary fields $D=1$. This allows to exactly renormalize the divergent part of the Hamiltonian density in 2 space dimensions, all variationally, without requiring any knowledge of the exact solution. With GCTNSs of moderate field bond dimension $D$, it is even possible to go near the gap closing, corresponding to the relativistic limit of the model we considered. Hence, GCTNSs have exactly the right UV properties to approximate non-relativistic QFTs and they can also accommodate relativistic theories provided the cutoff scale is only reasonably far from the physical mass scale (at least up to $\Lambda/m \sim 10^2$)

Naturally, interesting Hamiltonians are not quadratic and thus do not have a Gaussian ground state. In this context, a more general CTNS would be required, but it is worthwhile to see if GCTNSs can help already. For the Lieb-Liniger model, we showed GCTNSs allowed to obtain good approximations of the energy density, at least in the weak coupling regime. Importantly, the lowest bond field dimension $D=1$ already allows to capture essentially everything general Gaussian states (with infinitely many parameters) can capture. This means that GCTNSs allow a drastic compression of Gaussian states for quantum field Hamiltonians, yielding a potentially large gains for methods that build upon them. All these results are encouraging, and demonstrate that CTNSs indeed have the right properties expected from their discrete tensor network analogs, and, as a result, deserve to be studied further.

Promising extensions of this work are already possible while staying in the relatively easy realm of Gaussian states. The states we considered could be extended to deal with Fermions, where richer physics already appears for quadratic Hamiltonians. Dealing with multiple species of Bosons / Fermions could also enable the exploration of topological phases and see if their characterization for GCTNS matches what can be seen on the lattice. Further, in general, one can obtain much more from Gaussian states than mere ground states, and one could obtain the spectrum and real time dynamics with GCTNS extending the geometric methods developed in \cite{guaita2019,hackl2020}. Finally, the success of tensor network methods has been well understood from  their entanglement properties, and it would be useful to see if such an analysis can be done as well in the context of GCTNSs. In particular it is still unknown if the bond field dimension can upper bound the prefactor in the area law scaling of entanglement entropy, as it does in the discrete.

Going beyond the Gaussian setting to deal with genuine interacting theories could be done in different ways. A first step could be to stay with GCTNSs but considering a sum of them, which is no longer Gaussian. In this context, the fact that low field bond dimension and thus very few parameters give already good approximations of the best Gaussian states would allow to consider large sums. This would be prohibitively expensive in the more brutal approach of considering a sum of generic Gaussian states. GCTNSs could also be used to construct a better basis of states for Hamiltonian truncation methods. In this approach (see \eg \cite{rychkov2015}), one diagonalizes an interacting Hamiltonian in a truncated basis made from the low energy sector of the free Fock space. With GCTNSs, this free Fock space could be replaced by the Fock space built from the excitations above a GCTNSs optimized on the interacting Hamiltonian. 

Another, more radical option is to use genuinely non-Gaussian CTNSs. There, the difficulty is that it is not possible to compute correlation functions exactly, and in particular to compute the energy density one typically wants to minimize. As a first step, such correlations could be evaluated with Monte-Carlo or perturbatively. Note in the latter case we would still have an overall non-perturbative method: even at the lowest order of Taylor expansion, the Gaussian part would already contain non-perturbative effects in the model coupling constant. The most appealing option and the one also most in the spirit of tensor networks would be to evaluate correlation functions of a CTNS in $2$ dimensions using the transfer matrix method in $1$ dimension as proposed in \cite{tilloy2019}. In $1$ space dimension, one can use CMPSs to efficiently find the largest eigenstate of an operator, here the transfer matrix. This would reduce the problem of computing CTNS correlation functions to that of optimizing a CMPS. This would likely require an improvement of the efficiency of existing CMPS algorithms, but does not seem out of reach. Ultimately, although a lot remains to be done to make CTNSs practically useful in the context of interacting QFTs, we hope that the present work offers evidence that this is a path worth pursuing.

\vskip0.5cm
\noindent \textit{Note:} While finishing the present paper, we got aware of work conducted in parallel at the University of Ghent by Aelbrecht, Mortier, and Haegeman \footnote{The results are to appear in the master's dissertation of Bastiaan Aelbrecht.}.

%% file: _Appendix.tex
\section{Exact diagonalization of \texorpdfstring{$H$}{H}} \label{appendix:exact}

\noindent
Since it is quadratic, this Hamiltonian can be diagonalized exactly by Fourier and Bogoliubov transform.  The Fourier transform yields
\begin{equation}\label{eq:H_Fourier}
    H=\int\frac{\upd^d p}{(2\pi)^d}\left(p^2+\mu\right)\hpsi^{\dagger}_p\hpsi_p+\lambda\left(\hpsi^{\dagger}_p\hpsi^{\dagger}_{-p}+\hpsi_p\hpsi_{-p}\right).
\end{equation}
The Bogoliubov transform consists in introducing new creation annihilation operators $(\hb_p^\dagger,\hb_p)$ linearly related to the original ones
\begin{align}
    \hpsi_p&=u_p \hb_p + v_p\hb_{-p}^\dagger\\
    \hpsi^\dagger_p&=u^*_p \hb^\dagger_p + v^*_p\hb_{-p}\; ,
\end{align}
where $|u_p|^2-|v_p|^2=1$ to ensure the canonical commutation relations remain valid. The Hamiltonian \eqref{eq:H_Fourier} becomes diagonal if 
\begin{equation}
    u_p v_p=\frac{\lambda}{p^2+\mu}(u_p^2+v_p^2)\;,
\end{equation}
which is solved by
\begin{align}
u_p&=\sqrt{\frac{p^2+\mu}{2\sqrt{\left(p^2+\mu\right)^2-4\lambda^2}}+\frac{1}{2}}\, ,\\
v_p&=-\sqrt{\frac{p^2+\mu}{2\sqrt{\left(p^2+\mu\right)^2-4\lambda^2}}-\frac{1}{2}}\, .
\end{align}
Finally, the diagonalized Hamiltonian reads:
\begin{equation}
    H=\int\frac{\upd^d p}{(2\pi)^d} \; \varepsilon_1(p)\, \hb_p^\dagger \hb_p + \varepsilon_0(p),
\end{equation}
with
\begin{align}
    \varepsilon_1(p) &= \sqrt{(p^2+\mu)^2 - 4\lambda^2}\\
    \varepsilon_0(p) &=\frac{1}{2} \left[\varepsilon_1(p) - (p^2+\mu)\right]\,.
\end{align}
The associated ground state energy density $\mathbf{e}_0$, which will be useful for benchmarks, is
\begin{equation}
    \mathbf{e}_0= \int\frac{\upd^d p}{(2\pi)^d} \,\varepsilon_0(p)\, .
\end{equation}
For large $p$, $\varepsilon_0(p)$ decays as $p^{-2}$ and thus the ground energy density is infinite for $d\geq 2$.

Expressing $\hpsi_p$ as a function of the $\hb_p$, we get the ground state two point function:
\begin{equation}
    \langle \hpsi^\dagger_p \hpsi_q\rangle = \left(\frac{p^2+\mu}{2\, \varepsilon_1(p)} -\frac{1}{2}\right) \delta(p-q).
\end{equation}

\section{Correlation functions and their gradients} \label{appendix:correl}

\subsection{Two-point functions}
\noindent
The expectation values of the GCTNSs are computed as functional derivatives of the generating functional $\mathcal{Z}_{j',j}$ given in eq. \eqref{eq:generating_gaussian}. For example
\begin{align}
\langle\hpsi^\dagger(x) \hpsi(y)\rangle_{V,\alpha}
=& \frac{\delta}{\delta j'(x)}\frac{\delta}{\delta j(y)} \mathcal{Z}_{j',j} \bigg|_{j,j'=0}\,.
\end{align}
All the two point functions can be computed in the same way and we focus on this one for illustration. Using the expression for $\mathcal{Z}_{j',j}$ we get:
\begin{align} \label{eq:psidaggerpsi}
\langle \hpsi^{\dagger}(x)\hpsi(y)\rangle_{V,\alpha}=\int \frac{\upd^dp}{(2\pi)^d}   C_{V,\alpha}(p) e^{ip(x-y)}
\end{align}
with
\begin{align} \label{eq:C_split}
      C_{V,\alpha}(p)=(0,\alpha^{(1)*}) K(p)
(\alpha^{(1)},0)^T +\langle \hpsi^{\dagger}_p\rangle\langle\hpsi_p\rangle \delta(p),
\end{align}
where $(\alpha^{(1)},0)$ and $(0, \alpha^{(1)*})$ are $2D$ vectors, $K(p) = \left(p^2 \, \mathds{1} + W\right)^{-1}$ (see eq. \eqref{eq:W}) and only the second term (the zero mode) depends on $\alpha^{(0)}$ 
\begin{align} \label{eq:psi_0}
    \langle \hpsi_p^{\dagger} \rangle
    =
      \alpha^{(0)}[( \alpha^{(1)},\alpha^{(1)*})K(p) (\alpha^{(1)},0)^T]\, .
\end{align}
Note that in the models we are considering one can choose the gauge where $a^{(0)} \in \mathds{R}$. We set  \begin{equation}
    C_{V,\alpha}(p)\big|_{\alpha^{(0)}=0}=(0,\alpha^{(1)*}) K(p)
(\alpha^{(1)},0)^T
\end{equation} because it corresponds to the $C_{V,\alpha}$ when $\alpha^{(0)}=0$.

Let us now compute the real space correlation function at equal points, which is needed to get the energy density. As the momentum integral of the zero mode is trivial, we focus on the contribution of the contribution from $C_{V,\alpha}(p)\big|_{\alpha^{(0)}=0}$. First we diagonalize matrix $K(p)$ with an unitary $2D \times 2D$ matrix $U$, such that $W= U^{-1}L U$ and $L$ is a $2D \times 2D$ diagonal matrix with eigenvalues $\lambda_1,\lambda_2,...,\lambda_{2D}$. Note that the matrix $W$ needs to be positive definite for the state to be physical, thus $\re[\lambda_i] >0$. We get 
\begin{equation}
K(p)=U^{-1}(p^2\cdot \mathds{1} +L)^{-1}U,
\end{equation}
and hence
\begin{equation*}
    C_{V,\alpha}(p)\big|_{\alpha^{(0)}=0}=(0,\alpha^{(1)*}) U 
(p^2\cdot \mathds{1} +L)^{-1} U^{-1}  (\alpha^{(1)},0)^T\, .
\end{equation*}
This allows to find the equal point $2$-point function
\begin{align} \label{eq:psipsi_diagonalized}
\langle \hpsi^{\dagger}\!(x)\hpsi(x)\rangle|_{\alpha^{(0)}=0}\!=\!\!\sum_{i=1}^{2D}   [(0,\alpha^{(1)*}) U]_i I_1(\lambda_i) [U^{-1}  (\alpha^{(1)}\!,0)^T]_i 
\end{align}
with the integral 
\begin{equation*}
    I_1(\lambda_i)=\int \frac{\upd^dp}{(2\pi)^d}\frac{1}{p^2+\lambda_i}.
\end{equation*}
This integral is convergent in $d=1$ and logarithmically divergent in $d=2$. However, the divergences cancel each other in the sum \eqref{eq:psipsi_diagonalized} as they do not depend on $\lambda_i$ and thus the particle density is finite in $d=2$ space dimensions (and in fact even in $d=3$). 

We can proceed in the same way to compute the other correlation functions $\langle \hpsi(x)\hpsi(x)\rangle$ and $\langle \hpsi^\dagger(x)\hpsi^\dagger(x)\rangle$. For these, the logarithmic divergences do not cancel each other in $d=2$ and contribute to the divergence of the energy density we explain how to renormalize in the main text.

To compute the kinetic energy density, we simply take the derivative of the two point function $ \lim_{x \rightarrow y}\partial_x \partial_y \langle \hpsi^{\dagger}(x)\hpsi(y)\rangle$. Ultimately, this yields the same formula as before with the replacement of $I_1$ by
\begin{equation*}
    I_{1\text{kin}}(\lambda_i)=\int \frac{\upd^dp}{(2\pi)^d}\frac{p^2}{p^2+\lambda_i}. 
\end{equation*}
This latter integral is linearly divergent in $d=1$, but again this divergent part is independent of $\lambda_i$ and thus cancels in the expression for the kinetic energy. In $d=2$, the leading divergence is quadratic and cancels in the sum but a subleading logarithmic divergence subsists in the expression of the kinetic energy density, as well as in $\langle \hpsi(x)\hpsi(x)\rangle$ and $\langle \hpsi^\dagger(x)\hpsi^\dagger(x)\rangle$, contributing to the overall logarithmic divergence of the energy density.

\subsection{Four-point function} \label{sec:four_point}
\noindent
The Lieb-Liniger Hamiltonian contains quartic terms and, as a result, evaluating its energy density requires computing a 4-point function. As our states are Gaussian, we can use Wick's theorem or the expression for the generating functional $\mathcal{Z}_{j',j}$ to get:
\begin{align}\label{eq:4point}
    \langle \hpsi^{\dagger}\hpsi^{\dagger}\hpsi \hpsi \rangle=& |\langle\hpsi\rangle|^4\nonumber + 4 |\langle\hpsi\rangle|^2\, \langle \hpsi^{\dagger} \hpsi\rangle \mid_{\alpha^{(0)}=0} \nonumber\\
    &+ |\langle\hpsi\rangle|^2 \,\{ \langle \hpsi^{\dagger} \hpsi^{\dagger}\rangle +\langle\hpsi \hpsi \rangle\}\mid_{\alpha^{(0)}=0} 
 \nonumber \\
    &+2\langle\hpsi^{\dagger}\hpsi\rangle \mid_{\alpha^{(0)}=0} \langle \hpsi^{\dagger}  \hpsi\rangle \mid_{\alpha^{(0)}=0} \nonumber\\
    &+\langle\hpsi^{\dagger} \hpsi^{\dagger}\rangle\mid_{\alpha^{(0)}=0} \langle \hpsi \hpsi\rangle \mid_{\alpha^{(0)}=0},
\end{align}
where all the operators are taken in the same point $x$ which we omitted since the problem is translation invariant. We have also split the 2-point functions into a part that does not depend on $\alpha^{(0)}$ and the zero mode contribution:
\begin{equation}
\langle\hpsi\rangle:= \langle \hpsi(x)\rangle=\frac{1}{(2\pi)^d}\langle\hpsi_{p=0}\rangle\, .
\end{equation}
The latter corresponds to the condensed fraction in the Lieb-Liniger model. Taking the mean field approximation is equivalent to neglecting the last two lines in \eqref{eq:4point} as one assumes that the zero mode $\langle\hpsi\rangle$ dominates.

\subsection{Gradients}

\noindent
To carry the optimization we need the gradient of $\langle h\rangle_{V,\alpha}$ with respect to the $2D+1$ complex coefficients parameterizing the state ($D$ complex parameters from $M$, $D$ complex parameters from $\alpha^{(1)}$ vector and one parameter from $\alpha^{(0)}$).  
We present the computations for one 2-point function, $\langle\hpsi^\dagger(x)\hpsi(x)\rangle$, as the rest of the gradients are computed analogously. The derivative of the Fourier transformed 2-point function $C_{V,\alpha}(p)$  with respect to some GCTNS parameter $a$ is given by
\begin{align}
\frac{\partial C_{V,\alpha}(p)}{\partial a} =&\frac{\partial C_{V,\alpha}(p)}{\partial a}\bigg|_{\alpha^{(0)}=0} \nonumber \\
&+\langle \hpsi^{\dagger}_p\rangle\frac{\partial \langle \hpsi_p\rangle}{\partial a} \delta(p)+\frac{\partial \langle \hpsi^{\dagger}_p\rangle }{\partial a}\langle \hpsi_p\rangle \delta(p) \, .
\label{eq:der_correlation}
\end{align}
Let us compute the first term. Using the fact that \begin{equation*}
    \frac{d K^{-1}}{d x}= -K^{-1} \frac{d K}{dx} K^{-1}
\end{equation*} and
$ U^{-1}K^{-1} U=(p^2+L)^{-1}$ we obtain the derivative with respect to parameters of the mass matrix M:
\begin{align*} 
&\frac{\partial C_{V,\alpha}(p)}{\partial \re(m_j)/i\im(m_j)}\mid_{\alpha^{(0)}=0}\\
&=- (0,\alpha^{(1)*}) U 
(p^2 +L)^{-1} F^{R/I}(j) (p^2 +L)^{-1} U^{-1}  (\alpha^{(1)},0)^T
\end{align*}
where $F(j)$ is $2D\times2D$ a complex matrix with elements
\begin{align*}
    F_{lk}^{R}(j)&= U^{-1}_{lj}U_{jk}+U^{-1}_{l,j+D} U_{j+D,k}\\
    F^{I}_{lk}(j)&=U^{-1}_{lj}U_{jk}-U^{-1}_{l,j+D} U_{j+D,k}.
\end{align*}
The derivative with respect to the parameters of $\alpha^{(1)}$ gives:
\begin{align*} 
&\frac{\partial C_{V,\alpha}(p)}{\partial \re(\alpha_j)/ i\im(\alpha_j)}\mid_{\alpha^{(0)}=0}=
\\&= -(0,\alpha^{(1)*}) U (p^2+L)^{-1}  U^{-1} G^{R/I}(j)  U (p^2+L)^{-1}  U^{-1}  (\alpha^{(1)},0)^T\\
&+ [(0,\alpha^{(1)*}) U (p^2+L)^{-1}  U^{-1}]_j \pm  [ U (p^2+L)^{-1}  U^{-1}(\alpha^{(1)},0)]_{j+D}
\end{align*}
where $G^R(j)$ and $G^I(j)$ are $2D \times 2D$ matrices
\begin{align*}
    G^{R/I}(j)&= -[e_j \cdot (0,\alpha^{(1)*}) \pm e_{j+D}\cdot (\alpha^{(1)},0)\nonumber \\&+ (0,\alpha^{(1)*})^T\cdot e^T_{j} \pm(\alpha^{(1)},0)^T\cdot e^T_{j+D}]
\end{align*}
 with $e_j$ the $2D$ column vector with the $j$-th coefficient $1$ and zero otherwise. The precise form of these expressions is not crucial. In the end, what matters is that they contain two terms in $(p^2+L)^{-1}$. To go get the equal point correlation functions in real space, we need to integrate over momenta, which means we simply need to know the integral
 \begin{equation}
    I_2(\lambda_i, \lambda_j)=\int \frac{\upd^dp}{(2\pi)^d}\frac{1}{p^2+\lambda_i}\frac{1}{p^2+\lambda_j}\, ,
\end{equation}
which is well behaved in the dimensions we consider and given in appendix \ref{appendix:regulated}. The gradient of the kinetic term is obtained in the same way and brings a $p^2$ term in the previous integral
\begin{equation}
    I_{2\text{kin}}(\lambda_i, \lambda_j)=\int \frac{\upd^dp}{(2\pi)^d}\frac{p^2}{p^2+\lambda_i}\frac{1}{p^2+\lambda_j}\, .
\end{equation}
The zero mode terms in the gradient $\eqref{eq:der_correlation}$ are computed in the same way and the integration over $p$ is then immediate because of the Dirac $\delta$ in \eqref{eq:der_correlation} (only the zero mode contributes).

\section{A few (regulated) momentum integrals}\label{appendix:regulated}

\noindent
In one dimension,  we can compute $I_1(\lambda)$ with the theorem of residues or using the fact that $\arctan$ is an explicit primitive of the integrand to get:
\begin{align}
I_1(\lambda_i)=\int \frac{\upd p}{2\pi} \frac{1}{p^2+\lambda} \nonumber =\frac{1}{2\sqrt{\lambda}}.
\end{align}
Clearly, $I_{1\text{kin}} = \int \upd p/(2\pi) 1 - \lambda I_{1}$ and thus diverges. With a UV regulator $\Lambda$ (unrelated to the non-relativistic regulator of \ref{sec:modeldefinition}) we get:
 \begin{align*}
    I_{1\text{kin}}(\lambda)&=\int_{-\Lambda}^\Lambda \frac{\upd p}{2\pi} \frac{p^2}{p^2+\lambda}\\
    &=\frac{\Lambda}{\pi}-\frac{\sqrt{\lambda}}{\pi}\arctan(\Lambda/\sqrt{\lambda})\\
    &= \frac{\Lambda}{\pi}- \frac{\sqrt{\lambda}}{2} + o(1)
\end{align*}
The integrals $I_2(\lambda_i,\lambda_j)$ and $I_2(\lambda_i,\lambda_j)$ are convergent and computed with the theorem of residues which gives
 \begin{align*}
I_2(\lambda_i, \lambda_j)&=\int  \frac{\upd p}{2\pi} \frac{1}{p^2+\lambda_i} \frac{1}{p^2+\lambda_j}\\&=\frac{1}{2(\sqrt{\lambda_i}+\sqrt{\lambda_j})(\sqrt{\lambda_i}\sqrt{\lambda_j})} \,
\end{align*}
and
\begin{align*}
I_{2\text{kin}}(\lambda_i, \lambda_j)&= \int  \frac{\upd p}{2\pi} \frac{p^2}{p^2+\lambda_i} \frac{1}{p^2+\lambda_j} 
\\&=\frac{1}{2(\sqrt{\lambda_i}+\sqrt{\lambda_j})}\, .
\end{align*}
In two dimensions, $I_1$ already requires a UV regulator $\|p\|\leq \Lambda$
\begin{align*}
I_1(\lambda)&=\int_{\|p\|\leq \Lambda} \frac{\upd^2p}{(2\pi)^2} \frac{1}{p^2+\lambda} \nonumber \\
&=\int_0^{\Lambda^2} \frac{\upd(p^2)}{4\pi}\frac{1}{p^2+\lambda_i}\nonumber \\
&=\frac{1}{4\pi}\ln(\Lambda^2+\lambda)-\frac{1}{4\pi}\ln(\lambda)\\
&=\frac{1}{4\pi}\ln(\Lambda^2)-\frac{1}{4\pi}\ln(\lambda) +o(1)\, .
\end{align*}
Using the relation between $I_1$ and $I_{1\text{kin}}$ as in $d=1$ one gets
\begin{align*}
    I_{1\text{kin}}(\lambda)&=\int_{\|p\|\leq \Lambda} \frac{\upd^2p}{(2\pi)^2} \frac{p^2}{p^2+\lambda_i} \nonumber \\
  &=\frac{1}{4\pi}[\Lambda^2-\lambda \ln(\Lambda^2+\lambda)+\lambda \ln(\lambda)]\, .
\end{align*}
The integral $I_2$ is convergent in $d=2$ and computed with the theorem of residues
 \begin{align*}
I_2(\lambda_i, \lambda_j)&= \int  \frac{\upd^2p}{(2\pi)^2} \frac{1}{p^2+\lambda_i} \frac{1}{p^2+\lambda_j} \nonumber\\&=\begin{cases}
 \frac{\ln(\lambda_i/\lambda_j)}{4\pi (\lambda_i-\lambda_j)} , \text{for}\: \lambda_i \neq \lambda_j \\
 \frac{1}{4\pi \lambda_i}, \text{for} \:\lambda_i = \lambda_j 
\end{cases}\, .
\end{align*}
Finally, $I_{2\text{kin}}$ needs to be regulated. Using that formally $I_{2\text{kin}}(\lambda_i,\lambda_j) = I_1(\lambda_j)- \lambda_i I_2(\lambda_i,\lambda_j)$ we get
\begin{align*}
&I_{2\text{kin}}(\lambda_i, \lambda_j)=\int_{\|p\|\leq \Lambda} \frac{\upd^2p}{(2\pi)^2} \frac{p^2}{p^2+\lambda_i} \frac{1}{p^2+\lambda_j} \\
&= \frac{1}{4\pi}\ln(\Lambda^2)- \frac{\lambda_i\ln(\lambda_i) - \lambda_j\ln(\lambda_j)}{4\pi (\lambda_i-\lambda_j)} + o(1) \,.
\end{align*}